\documentclass[manuscript]{aastex} 

\shorttitle{Testing the Distance-Duality Relation with Galaxy Clusters and Type Ia Supernovae} 
\shortauthors{Djorgovski et al.} 

\begin{document} 


\title{Testing the Distance-Duality Relation with Galaxy Clusters and Type Ia Supernovae} 


\author{R. F. L. Holanda and J. A. S. Lima} 
\affil{Departamento de Astronomia, Universidade de S\~ao Paulo--USP, 
\\05508-900 S\~ao Paulo, SP, Brazil} 

\author{M. B. Ribeiro} 
\affil{Instituto de F\'isica, Universidade Federal do Rio de 
Janeiro--UFRJ, 
\\Rio de Janeiro, RJ, Brazil} 
\email{limajas@astro.iag.usp.br} 



\begin{abstract} 
In this letter we propose a new and model-independent cosmological 
test for the distance-duality (DD) relation, 
$\eta=D_{L}(z)(1+z)^{-2}/D_{A}(z)=1$, where $D_{L}$ and $D_{A}$ are, 
respectively, the luminosity and angular diameter distances. For 
$D_L$ we consider two sub-samples of SNe type Ia taken from 
Constitution data (2009) whereas $D_A$ distances are provided by two 
samples of galaxy clusters compiled by De Fillipis et al. (2005) and 
Bonamente et al. (2006) by combining Sunyaev-Zeldovich effect (SZE) 
and X-ray surface brightness.  The SNe Ia redshifts of each 
sub-sample were carefully chosen to coincide with the ones of the 
associated galaxy cluster sample ($\Delta z<0.005$) thereby allowing 
a direct test of DD relation. Since for very low redshifts, 
$D_{A}(z) \approxeq D_{L}(z)$, we have tested the DD relation by 
assuming that $\eta$ is a function of the redshift parametrized by 
two different expressions: $\eta(z) = 1 + \eta_{0}z$ and $\eta(z) = 
1 + \eta_{0}z/(1+z)$, where $\eta_0$ is a constant parameter 
quantifying a possible departure from the strict validity of the 
reciprocity relation ($\eta_0=0$). In the best scenario (linear 
parametrization) we obtain  $  \eta_{0} = -0.28^{+ 0.44}_{- 0.44}$ 
 ($2\sigma$, statistical + systematic errors)  for de Fillipis et al. sample (elliptical geometry), a 
result only marginally compatible with the DD relation. However, for 
Bonamente et al. sample (spherical geometry) the constraint is 
$  \eta_{0} = -0.42^{+ 0.34}_{- 0.34}$ ($3\sigma$, statistical + systematic errors) which is clearly 
incompatible with the duality-distance relation.\end{abstract} 





\section{Introduction} 

The Etherington's reciprocity relation (Etherington 1933) is of 
fundamental importance in cosmology. Its most useful version in the 
astronomical context, sometimes referred as \textit{distance-duality 
relation}, relates the \textit{luminosity distance} $D_{\scriptstyle 
L}$ with the \textit{angular diameter distance} $D_{\scriptstyle A}$ 
by means of the following expression, 
\begin{equation} 
 \frac{D_{\scriptstyle L}}{D_{\scriptstyle A}}{(1+z)}^{-2}=1. 
 \label{rec1} 
\end{equation} 
This equation is \textit{completely general}, valid for \textit{all} 
cosmological models based on Riemannian geometry, being dependent 
neither on Einstein field equations nor the nature of matter-energy 
content. It only requires that source and observer are connected by 
null geodesics in a Riemannian spacetime and that the number of 
photons are conserved. Therefore, it is valid for spatially 
homogeneous and isotropic (anisotropic) cosmologies, as well as for 
inhomogeneous cosmological models (Ellis 2007). 

The distance-duality (DD) relation plays an essential role in 
modern cosmology, ranging 
from gravitational lensing studies (Schneider et al. 1999) to analyzes from galaxy clusters observations (Cunha et al. 
2007, Mantz et al. 2009), as well as the plethora of cosmic 
consequences from primary and secondary temperature anisotropies of 
the cosmic microwave blackbody radiation (CMBR) observations 
(Komatsu et al. 2010). Even the the temperature shift equation 
$T_o=T_e/(1+z)$, where $T_o$ is the observed temperature and $T_e$ 
is the emitted temperature, a key result for analyzing CMBR 
observations, as well as the optical theorem that surface brightness 
of an extended source does not depend on the angular diameter 
distance of the observer from the source (an important result for 
understanding lensing brightness), are both consequences of 
Etherington' reciprocity relation (Ellis 1971, 2007). 

The Etherington law, as it is also sometimes called, has so far been 
taken for granted by virtually all analyzes of cosmological 
observations. Despite this, the distance-duality relation is in 
principle testable by means of astronomical observations. If one is 
able to find cosmological sources whose intrinsic luminosities are 
known (standard candles) as well as their intrinsic sizes (standard 
rulers), one can determine both $D_{\scriptstyle L}$ and 
$D_{\scriptstyle A}$, and after measuring the common redshifts, to 
test directly the above Etherington's result. Note that ideally both 
quantities must be measured in a way that does not utilize any 
relationship coming from a cosmological model, that is, they must be 
determined by means of intrinsic astrophysically measured 
quantities. 

The method described above for testing the reciprocity law is very 
difficult to carry out in practice due to limitations in our current 
understanding of galaxy evolution and, hence, one must still rely on 
less than ideal methods for seeking observational falsification of 
the reciprocity law. These less-than-ideal methods usually assume a 
cosmological model suggested by a set of observations, apply this 
model in the context of some astrophysical effect and attempt to see 
if the reciprocity relation remains valid. In this way, Uzan, 
Aghanim \& Mellier (2004) showed that observations from 
Sunyaev-Zeldovich effect (SZE) and X-ray surface brightness from 
galaxy clusters  offer a test for the distance-duality relation. It 
was argued that the SZE + X-ray technique  for measuring the angular 
diameter distances (ADD) (Sunyaev and Zeldovich 1972, Cavaliere and 
Fusco-Fermiano 1978) is strongly dependent on the validity of this 
relation. When the relation does not hold, the ADD determined from 
observations is $D^{cluster}_{A}(z)=D_{A}(z)\eta^{2}$ (actually, 
multiplied by $\eta^{-2}$ in their notation). Such a quantity 
reduces to the angular diameter distance only when the reciprocity 
relation is strictly valid, i.e., when $\eta=1$. They considered 18 
ADD galaxy clusters from Reese { et al.} sample (2002) for which 
a spherically symmetric cluster geometry has been assumed. Their 
analysis carried out in a $\Lambda$CDM model (Spergel { et al.} 
2003) shows that no violation of the distance-duality is only 
marginally consistent. 

Later on, De Bernardis, Giusarma \& Melchiorri (2006) also searched 
for deviations from the distance-duality relation by using the ADD 
from galaxy clusters provided by the sample of Bonamente {\it et 
al.} (2006). They obtained  a non violation of the distance-duality 
in the framework of the cosmic concordance $\Lambda$CDM model. 
Recently, Avgoustidis {et al.} (2010) used the distance 
relation, $d_L=d_A(1+z)^{2+\epsilon}$, in a flat $\Lambda$CDM model 
for constraining the cosmic opacity by combining recent SN Type Ia 
data compilation (Kowalski { et al.} 2008) with the latest 
measurements of the Hubble expansion at redshifts on the range $0 < 
z < 2$ (Stern { et al.} 2010). They found 
$\epsilon=-0.04_{-0.07}^{+0.08}$ (2$\sigma$). However, what was 
really being tested in the quoted works  was the consistency between 
the assumed cosmological model and some results provided by a chosen 
set of astrophysical phenomena. 

Following another route, Holanda, Lima \& Ribeiro (2010) discussed 
the consistency between the strict validity of the distance-duality 
relation and the assumptions about the geometry, elliptical and 
spherical $\beta$ models, used to describe the galaxy clusters. They 
used the function $\eta(z)$ parametrized in two distinct forms, 
$\eta = 1+\eta_{0}z$ and $\eta = 1+\eta_{0}z/(1+z)$, thereby 
recovering the equality between distances only for very low 
redshifts, in order to test possible deviations. By comparing the De 
Filippis {\it et al.} (2005) (elliptical $\beta$ model) and 
Bonamente {\it et al.} (2006) (spherical $\beta$ model) samples with 
theoretical $D^{Th}_{A}$ obtained from $\Lambda$CDM (Komatsu {\it et 
al.} 2010), they showed that the elliptical geometry is more 
consistent ($\eta_{0}=0$ in 1$\sigma $) with no violation of the 
distance-duality relation in the context of $\Lambda$CDM (WMAP7). 

The possibility to test new physics with basis on the validity of DD 
relation was first discussed by Basset \& Kuns (2004). They used 
current supernovae Ia data as measurements of $D_{\scriptstyle L}$ 
and estimated $D_{\scriptstyle A}$ from FRIIb radio galaxies (Daly 
\& Djorgovski 2003) and ultra compact radio sources (Gurvitz 1994, 
1999; Lima \& Alcaniz 2000, 2002, Santos \& Lima 2008). A moderate 
violation ($2\sigma$) caused by the brightening excess of SNe Ia at 
$z>0.5$ was found. In the same vein, De Bernardis, Giusarma \& 
Melchiorri (2006) also compared the ADD from galaxy clusters with 
luminosity distance data from supernovae to obtain a model 
independent test. In order to compare the data sets they considered 
the weighted average of the data in 7 bins and found that $\eta = 1 
$ is consistent in $68\%$ confidence level (1$\sigma$). However, one 
needs to be careful when using SZE + X-ray technique for measuring 
angular diameter distances to test the DD relation because such a 
technique is also dependent of its validity. In fact, when the 
relation does not hold, the ADD determined from observations is in 
general $D^{cluster}_{A}(z)=D_{A}(z)\eta^{2}$, which reduces to 
$D_{A}$ only if $\eta =1$. So, their work did not test DD relation, 
at least not in a consistent way. In addition, both authors binned 
their data, and, as such,  their results may have been influenced by 
the particular choice of redshift binning. 

In this context, the aim of this paper is to propose a consistent 
cosmological-model-independent test for equation (\ref{rec1}) by 
using two sub-samples of SNe Ia chosen from Constitution data (M. 
Hicken {  et al.} 2009) and two angular diameter distance (ADD) 
samples from galaxy clusters obtained through Sunyaev-Zeldovich 
effect and X-ray measurements with different assumptions concerning 
the geometry used to describe the clusters: elliptical $\beta$ model 
and spherical $\beta$ model. Following Holanda, Lima \& Ribeiro 
(2010), our analysis here will be based on two parametric 
representations for a possible redshift dependence of the distance 
duality expression, namely, 
\begin{equation} 
 \frac{D_{\scriptstyle L}}{D_{\scriptstyle A}}{(1+z)}^{-2}= \eta(z), 
 \label{rec} 
\end{equation} 
where 

\hspace{1.0cm} I. $\eta (z) = 1 + \eta_{0} z$, 

\hspace{1.0cm} II. $\eta (z) = 1 + \eta_{0}z/(1+z)$.\\ 

\noindent For a given pair of data set (SNe Ia, galaxy clusters), 
one should expect a likelihood of $\eta_0$ peaked at $\eta_0=0$, in 
order to satisfy the DD relation. It is also worth noticing that in 
our approach the data do not need to be binned  as assumed in some 
analyses involving DD  relation. As we shall see, for the Bonamente 
et al. sample (2006), where a spherical geometry was assumed, our 
results show a strong violation ($>3\sigma$) DD relation when the 
SNe Ia and galaxy clusters data are confronted. However, when the 
elliptical geometry is assumed (de Fillipis et al. 2005), the 
results are marginally compatible within $2\sigma$ with the distance 
duality relation. 

\section{Samples }\label{sec:Sample} 

In order to constrain the possible values of $\eta_{0}$ let us now 
consider two samples of ADD from galaxy clusters obtained by 
combining their SZE and X-ray surface brightness observations. The 
first one is formed by 25 galaxy clusters from De Filippis {{et 
al.}} (2005) sample. Since {\it Chandra } and {\it XMM} 
observations of clusters in the past few years have shown that in 
general clusters exhibit elliptical surface brightness maps, De 
Filippis et al. (2005) studied and corrected, using an isothermal 
elliptical $\beta$ model to describe the clusters, the ${\cal{D}}_A$ 
measurements for two samples for which combined X-ray and SZE 
analysis has already been reported using a isothermal spherical 
$\beta$ model. One of the samples, compiled by Reese {\it{et al.}} 
(2002), is a selection of 18 galaxy clusters distributed over the 
redshift interval $0.14 < z < 0.8$. The other one, the sample of 
Mason {\it{et al.}} (2001), has 7 clusters from the X-ray limited 
flux sample of Ebeling {\it{et al.}} (1996). The second is defined 
by the 38 ADD galaxy clusters from Bonamente { et al.} (2006) 
sample, where the cluster plasma and dark matter distributions were 
analyzed assuming hydrostatic equilibrium model and  spherical 
symmetry, thereby accounting for radial variations in density, 
temperature and abundance. This sample consists of clusters 
that have both X-ray data from the {\it Chandra Observatory} and SZE 
data from the BIMA/OVRO SZE imaging project, which uses the 
Berkeley-Illinois-Maryland Association (BIMA) and Owens Valley radio 
observatory (OVRO) interferometers to image  the SZE For the 
luminosity distances, we choose two  sub-samples of SNe Ia from 
Constitution SNe Ia data set whose redshifts coincide with the ones 
appearing in the galaxy cluster samples. In Fig. (1a) we plot $D_A$ 
multiplied by $(1+z)^{2}$ from galaxy clusters sample compiled by De 
Filippis { et al.} (2005) { (the errors bars contain statistical 
and systematic contributions)} and $D_{L}$ from our first SNe Ia 
sub-sample. In Fig. (1b) we plot the subtraction of redshift between 
clusters and SNe Ia. We see that the biggest difference is $\Delta z 
\approx 0.01$ for 3 clusters (open squares) while for the remaining 
22 clusters we have $\Delta z < 0.005$. In order to avoid the 
corresponding bias, the 3 clusters will be removed from all the 
analyzes presented here so that $\Delta z < 0.005$ for all pairs. 

Similarly, in Fig. (2a) we plot $D_A$ multiplied by $(1+z)^{2}$, but 
now for the Bonamente { et al.} (2006) sample { (errors bars also 
include statistical and systematic contributions)} and $D_{L}$ from 
our second SNe Ia sub-sample. In Fig. (2b) we display the redshift 
subtraction between clusters and SNe Ia. Again, we see that for 35 
clusters $\Delta z < 0.005$. The biggest difference is $\Delta z 
\approx 0.01$ also for 3 clusters, and, for consistency, they will 
also be removed from our analysis (next section). 

\section{Analysis and Results}\label{sec:analysis } 

Let us now estimate the  $\eta_{0}$ parameter for each sample in 
both parametrization for $\eta(z)=D_{L}(z)(1+z)^{-2}/D_{A}(z)$, 
namely, $\eta(z) = 1+\eta_{0}z$ and $\eta(z) = 1+\eta_{0}z/(1+z)$. 
It should be stressed that in general the SZE + X-ray surface 
brightness observations technique do not give $D_{A}(z)$, but 
$D^{cluster}_{A}(z)=D_{A}(z)\eta^{2}$. So, if one wishes to test 
equation (\ref{rec1}) with  SZE + X-ray observations from galaxy 
clusters, the angular diameter distance $D_{A}(z)$ must be replaced 
by $D^{cluster}_{A}(z)\eta^{-2}$ in equation (2). In this way, we 
have access to $\eta(z)=D^{cluster}_{A}(z)(1+z)^{2}/D_{L}(z)$. 

Following standard lines, the likelihood  estimator is determined by 
$\chi^{2}$ statistics 
\begin{equation} 
\label{chi2} \chi^{2} = \sum_{z}\frac{{\left[\eta(z) - \eta_{obs}(z) 
\right] }^{2}}{\sigma^2_{\eta_{obs}} }, 
\end{equation} 
where $\eta_{obs}(z) = (1+z)^{2}D^{cluster}_{A}(z)/D_{L}(z)$ and 
$\sigma^2_{\eta_{obs}}$ are the errors associated with the 
observational techniques.  For the galaxy cluster samples the 
common statistical contributions are: SZE point sources $\pm 8$\%, 
X-ray background $\pm 2$\%, Galactic N$_{H}$ $\leq \pm 1\%$, $\pm 
15$\% for cluster asphericity, $\pm 8$\% kinetic SZ and for CMB 
anisotropy $\leq \pm 2\%$. Estimates for systematic effects are as 
follow: SZ calibration $\pm 8$\%, X-ray flux calibration $\pm 5$\%, 
radio halos $+3$\% and X-ray temperatute calibration $\pm 7.5$\%. We 
stress that typical statistical errors amounts for nearly $20$\% in 
agreement with other works (Mason et al. 2001; Reese et al. 2002, 
Reese 2004), while for systematics we also find typical errors around + 
12.4\% and - 12\%  (see also table 3 in Bonamente { et al.} 2006). 
In the present analysis we have combined the statistical and 
systematic errors in quadrature for the angular diameter distance 
from galaxy clusters (D'Agostino 2004). 

On the other hand, after nearly 500 Ia  SNe discovered, the 
constraints on the cosmic parameters from luminosity distance are 
now limited by systematics rather than by statistical errors. In 
principle, there are two main sources of systematic uncertainty in 
SNe cosmology which are closely related to photometry and possible 
corrections for light-curve shape (Hicken at al 2009). However, at 
the moment it is not so neat how to estimate the  overall systematic 
effects for this kind of standard candles (Komatsu et al. 2010), 
and, therefore, we will neglect them in the following analysis. The 
basic reason is that systematic effects from galaxy clusters seems 
to be larger than  the ones of SNe observations, but their inclusion 
do not affect appreciably the results concerning the validity of the 
distance-duality relation. 

In figures (3a) and (3b) we plot the likelihood distribution 
function for each sample. For De Filippis { et al.}  we obtain 
$ { \eta_{0} = -0.28^{+ 0.44}_{- 0.44}}$  ($ {\chi_{d.o.f.}^2 = 1.02}$) and 
${\eta_{0} =-0.43^{+ 0.6}_{- 0.6}}$  (${ \chi_{d.o.f.}^2 = 1.03}$) in 
{ 2$\sigma$ (statistical + systematic errors)}. For Bonamente { et al.} we obtain   ${ \eta_{0} = 
-0.42^{+ 0.34}_{- 0.34}}$  (${ \chi_{d.o.f.}^2 = 0.88}$) and  $ { \eta_{0} = 
-0.66^{+ 0.5}_{- 0.5}}$  ($ { \chi_{d.o.f.}^2 = 0.86}$) in { 3$\sigma$ (statistical + systematic errors)}. We 
can see that the confrontation between the ADD from the former 
sample with SNe Ia data, points to a moderate violation of the 
reciprocity relation (the DD relation is marginally satisfied in 
$2\sigma$). This result remains valid even when only clusters with 
$z>0.1$ are considered. In this case we obtain  ${ \eta_{0} = -0.29^{+ 
0.34}_{- 0.34}}$ (${\chi_{d.o.f.}^2 = 0.91}$) within { 2$\sigma$ (statistical + systematic errors)}. However, 
for the Bonamente et al. sample, where a spherical $\beta$ model was 
assumed to describe the clusters, we see that DD relation is not 
obeyed even at 3$\sigma$. 

\section{Conclusions}\label{sec:Conclusions} 

In this letter, we have discussed a new and model-independent 
cosmological test for the distance-duality relation, $\eta(z) = 
D_{L}(1+z)^{-2}/D_{A}$. The basic idea of our statistical test is 
very simple. We consider the angular diameter distances from galaxy 
clusters (two samples) which are obtained by using SZE and X-ray 
surface brightness together the luminosity distances given by two 
sub-samples of SNe Ia taken from the Constitution data. The key 
aspect is that the SNe Ia sub-samples were carefully chosen in order 
to have the same redshifts of the galaxy clusters ($\Delta z< 
0.005$). For the sake of generality, the $\eta(z)$ parameter was 
also parameterized in two distinct forms, namely, $\eta = 
1+\eta_{0}z$ and $\eta = 1+\eta_{0}z/(1+z)$, thereby recovering the 
equality between distances only for very low redshifts. It should be 
noticed that in our method the data do not need to be binned. { 
Interestingly, although independent of any cosmological scenario, 
our analysis depends on the starting physical hypotheses describing 
the galaxy clusters.} 

By comparing the De Filippis { et al.} (2005) (elliptical $\beta$ 
model) and Bonamente { et al.} (2006) (spherical $\beta$ model) 
samples with two sub-samples of SNe Ia, we show that the elliptical 
geometry is more consistent with no violation of the 
distance-duality relation. In the case of De Filippis { et al.} 
(2005) sample (see Fig. 3a) we find  ${ \eta_{0} = - 0.28^{+ 0.44}_{- 
0.44}}$ and   ${\eta_{0} = -0.43 ^{+ 0.6}_{- 0.6}}$  for linear and 
non-linear parametrization in 2$\sigma$  { (statistical + systematic errors)}, respectively. On the other 
hand, the spherical $\beta$ model (see Fig. 3b) is not compatible 
with the validity of the distance-duality relation. For this case we 
obtain  $ \eta_{0} = -0.42^{+ 0.34}_{- 0.34}$  and  ${\eta_{0} = -0.66^{+ 
0.5}_{- 0.5}}$  for linear and non-linear parameterizations in 
3$\sigma$ {  (statistical + systematic errors)}, respectively. 

Finally, it is also interesting to compare the present results with 
the ones of Holanda, Lima \& Ribeiro (2010). Their analysis revealed 
that the isothermal elliptical $\beta$ model is compatible with the 
Etherington theorem at 1$\sigma$ moduli the $\Lambda$CDM model while 
the non isothermal spherical model is only marginally compatible at 
3$\sigma$. Here as there, the sphericity assumption for the cluster 
geometry resulted in a larger incompatibility with the validity of 
the duality relation in comparison with an isothermal non spherical 
cluster geometry.


\clearpage 
\begin{figure}[h!] 
   \centering 
       \includegraphics[width=0.45\linewidth]{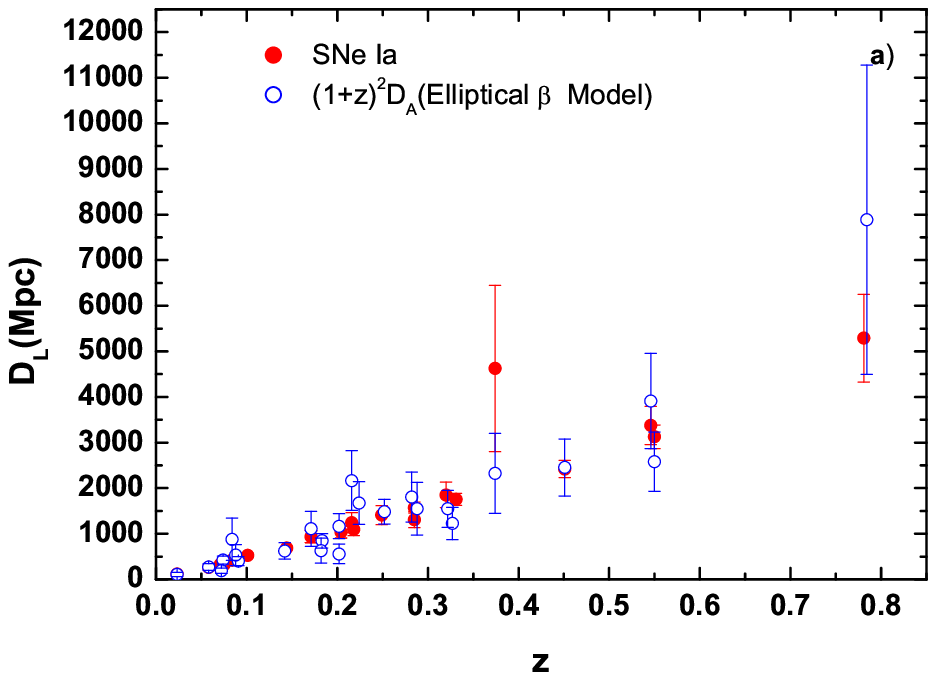} 
       \includegraphics[width=0.45\linewidth]{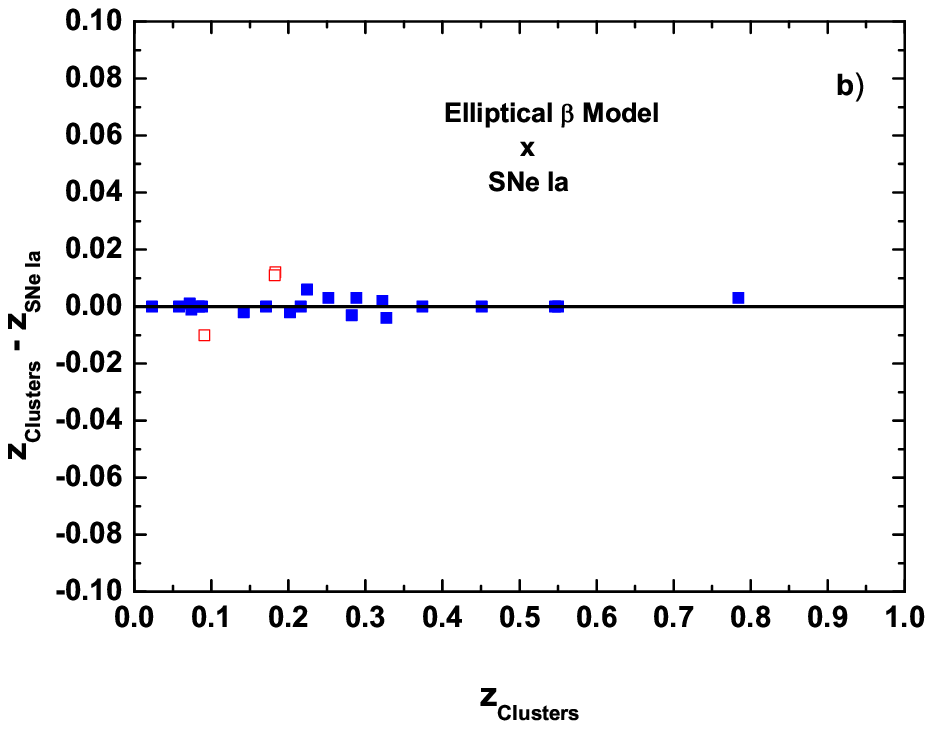} 
   \caption{a) Galaxy clusters and SNe Ia data. The open (blue) 
   and filled (red) circles with the associated error bars stand, 
   respectively, for the De Filippis et al. (2005) and SNe Ia 
   samples. b) The redshift subtraction for the same pair of cluster-SNe Ia samples. 
   The open squares represent the pairs of points for which $\Delta z \approx 0.01$.} 
\end{figure} 

\begin{figure}[h!] 
   \centering 
       \includegraphics[width=0.45\linewidth]{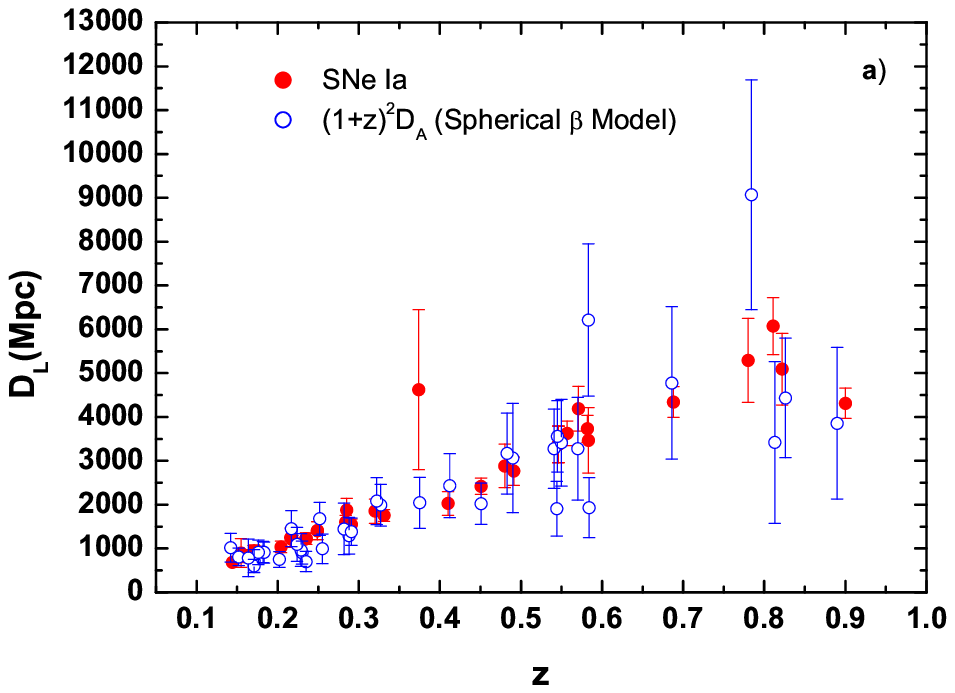} 
       \includegraphics[width=0.45\linewidth]{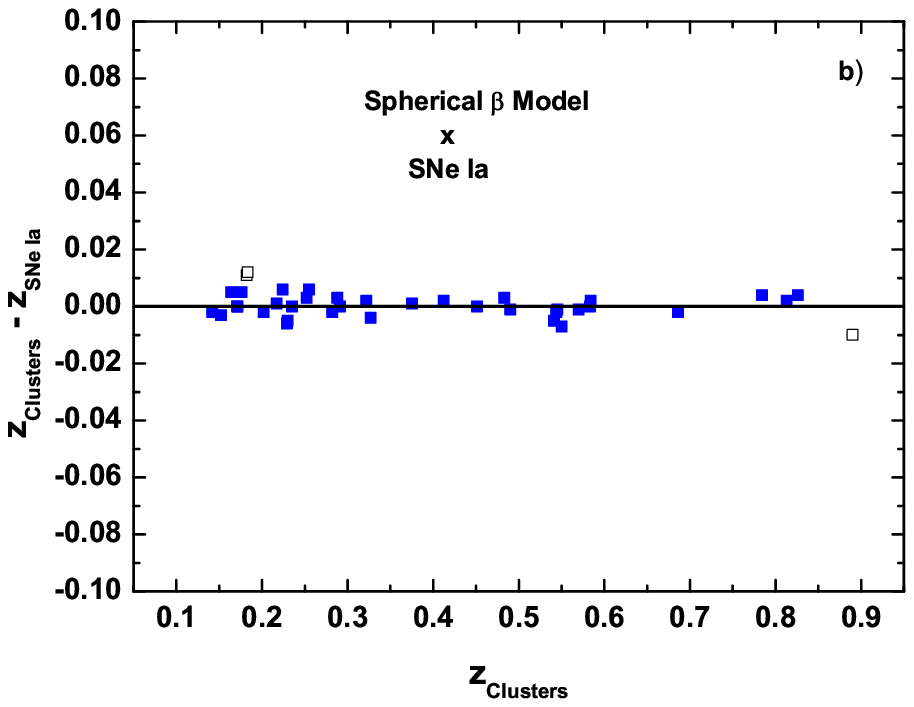} 
   \caption{a) Galaxy clusters and SNe Ia data. The open (blue) 
   and filled (red) circles with the associated error bars stand, 
   respectively, for the Bonamente et al. (2006) and SNe Ia samples. 
   b) The redshift subtraction for the same pair of cluster-SNe Ia samples. 
   As in Fig. 1b, the open squares represent the pairs of points with the biggest difference in redshifts ($\Delta z \approx 0.01$).} 
\end{figure} 

\begin{figure}[h!] 
{\includegraphics[width=85mm, angle=0]{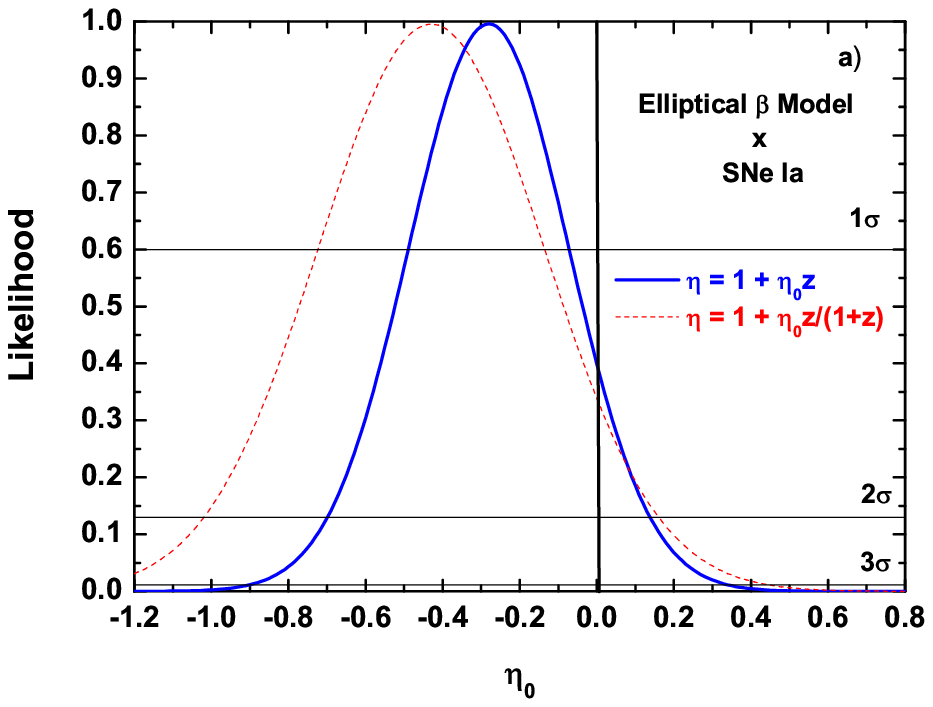} 
\includegraphics[width=85mm, angle=0]{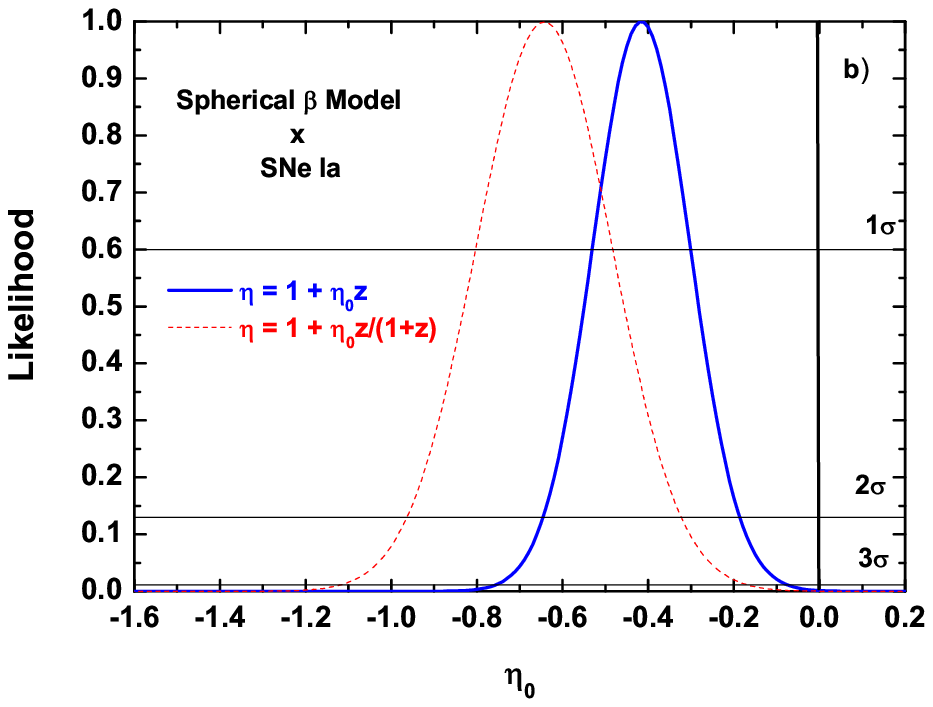} 
\hskip 0.1in} \caption{ { {a)}} The likelihood distribution 
functions for De Filippis { et al.} sample for both 
parametrizations. { {b)}} The likelihood distribution functions 
for  Bonamente { et al.} sample. Note that the elliptical model 
is compatible with the Etherington theorem at 2$\sigma$ while the 
spherical model is not compatible.} \label{fig:Analysis} 
\end{figure} 



\end{document}